\documentclass[11pt,aps,prd,preprint,superscriptaddress]{revtex4-1}
\usepackage{amsmath}
\usepackage[dvips]{graphicx}
\usepackage[english]{babel}
\usepackage{epsfig}

\begin{document}
\title{Lorentz Invariance Violation induced time delays in GRBs in different cosmological models}
\author{M. Biesiada\footnote{E-mail: biesiada@us.edu.pl} and A. Pi{\'o}rkowska\footnote{E-mail: apiorko@us.edu.pl}}
\affiliation{Department of Astrophysics and Cosmology, Institute of Physics, University of Silesia,\\ Uniwersytecka 4, 40-007 Katowice, Poland.}

\begin{abstract}

Lorentz Invariance Violation (LIV) manifesting itself by energy dependent modification of standard relativistic dispersion relation has recently attracted a considerable attention. Ellis et al. previously investigated the energy dependent time offsets in different energy bands on a sample of gamma ray bursts and, assuming standard cosmological model, they found a weak indication for redshift dependence of time delays suggestive of LIV.

Going beyond the $\Lambda$CDM cosmology we extend this analysis considering also four alternative models of dark energy (quintessence with constant and variable equation of state, Chaplygin gas and brane-world cosmology). It turns out that the effect noticed by Ellis et al. is also present in those models and is the strongest for quintessence with variable equation of state.

\end{abstract}

\maketitle

\section{Introduction}

Modern approaches to quantum gravity predict Lorentz Invariance Violation (LIV) as a consequence of the space-time foamy structure at small scales. Such violation can manifest itself by energy dependent modification of standard relativistic dispersion relation \cite{Amelino-Camelia,Mk 501}.

Several years ago Amelino-Camelia et al. \cite{Amelino-Camelia} proposed to use astrophysical objects to look for energy dependent time of arrival delays. Specifically gamma ray bursts (GRBs) -- very high energetic events visible from cosmological distances -- are the most promising sources of constraining LIV theories \cite{Ellis1,Ellis2,Rodrigues,Piran Jacob}. Recently Zhuk et al \cite{Zhuk} have also used GRB bounds on LIV due to extra dimensions.

The idea of searching for time of flight delays is tempered however by our ignorance concerning intrinsic delay (at  the source frame) in different energy channels (see e.g. \cite{Ellis2}). This clearly disfavors using energy dependent patterns in time-of-flights from single sources. As a possible way out of this trouble (and a first step in disentangling intrinsic time delays) Ellis et al. \cite{Ellis2} proposed to work on a statistical ensamble of GRBs and formulated the problem in terms of linear regression where the intercept represents intrinsic time delay and the linear term represents LIV effect. They found a weak evidence for LIV (in \cite{Ellis2} further corrected in \cite{Piran Errata}). On the other hand it was also shown \cite{Biesiada Piorkowska}, that lack of detailed knowledge about cosmological model (in the context of accelerating expansion of the Universe) could be another source of systematic effects at high redshifts. Moreover, despite calling $\Lambda$CDM scenario ``the concordance model'' there exists voices -- the most recent belonging to Linder \cite{Linder} -- to consider several alternative scenarios such like the quintessence (scalar field with evolving equation of state) on equal footing in as many cosmological tests (CMBR, LSS, lensing etc.) as possible and only then judge which one is better supported by the data (see also \cite{Rubin,BiesiadaJCAP,Wood-Vasey}).

This suggests to ask how does such fitting of time delays vs. certain (defined below) function of cosmic expansion rate look in those alternative models. Or maybe the effect noticed by Ellis et al. could be attributed to incorrectly assuming the $\Lambda$CDM background whereas the real Universe's geometry is more accurately described by some other model (like the quintessence for example) taken from the inventory of accelerating cosmologies. Perhaps the redshift dependence of time delays vanishes in some other realistic (i.e. non-excluded by cosmological tests) models of presently accelerating Universe. This is the main idea behind the present paper.

\section{LIV induced time delays in cosmological sources}

Let us consider a phenomenological approach for LIV  by assuming (after \cite{Mk 501} for better comparision of results) the modified dispersion relation for photons in the form:
\begin{equation}
E^2 - p^2 c^2 = \epsilon E^2 \left( \frac{E}{{\xi}_n
E_{QG}}\right)^n \label{dispersion}
\end{equation}
where:$\epsilon = \pm 1$ is ``sign parameter'' \cite{Mk 501}, ${\xi}_n$ is a dimensionless parameter. One may assume (as a first guess) $E_{QG}$ equal to the Planck energy, ${\xi}_1 =1$ and ${\xi}_2=10^{-7}$ \cite{Piran Jacob}. The dispersion relation (\ref{dispersion}) essentially corresponds to the power-law expansion (see \cite{Ellis1}) so for practical purposes (due to smallness of expansion parameter $E/E_{QG}$) only the lowest terms of the expansion are relevant.
Because in some LIV theories the odd power terms might be forbidden \cite{Burgess} usually the cases of $n=1$ and $n=2$ are retained. For sake of being comparable with the results of Ellis et al. \cite{Piran Errata} and because the LIV effects are small for sources at low and moderate redshifts used in this study, we retain only the $n=1$ term, and moreover we assume the sign parameter $\epsilon = +1$ in
the formulae.

Time of flight for the photon of energy $E$ is equal to \cite{Piran Jacob,Piran Errata,Biesiada Piorkowska}
\begin{equation}
t_{LIV} = \int_{0}^{z} \lbrack 1 +   \frac{E}{E_{QG}} (1+z')
\rbrack \frac{dz'}{H(z')} \label{flight time}
\end{equation}
Consequently, the time delay between a low energy and a high energy photon with the energy difference $\Delta E$ takes the following form
\begin{equation}
\Delta t_{LIV} =    \frac{\Delta E}{E_{QG}}
  \int_0^z  \frac{(1+z')dz'}{H(z')} \label{time delay}
\end{equation}
where: $H(z) = H_0 h(z)$ is cosmological expansion rate (the Hubble function).

The observational strategy emerging form (\ref{time delay}) is very simple: monitor appropriate (i.e. emitting both low and high energy photons) cosmological source at different energy channels and try to detect this time delay. Following this line Ellis et al. (\cite{Ellis1,Ellis2,Ellis new}) used a sample of gamma-ray bursts (GRBs) with known redshifts.
However there remains an indispensable uncertainty: there is no reason for which low and high energy signal should be emitted simultaneously, and while detecting distinct signals (peaks in the light curve) at different energies we have no idea which one was sent first. This is known as so-called intrinsic time lags problem.

In \cite{Ellis2} it has been noticed that while this ambiguity clearly disfavors using energy dependent patterns in time-of-flights from single sources, one is still able to search for statistical correlations of spectral time lags with redshift in an ensamble of sources located at different redshifts. If one
decomposes the observed time delay $\Delta t_{obs}$ between different energy channels:
$\Delta t_{obs} = \Delta t_{LIV} + \Delta t_{intrinsic}$ then, using the notation of \cite{Ellis2}, after taking into account cosmological time dilation factor $1+z$, one has $\Delta t_{obs} = a_{LIV} (1+z) K + b (1+z)$, where: $K = \frac{1}{1+z} \int^z_0 \frac{(1+z')dz'}{h(z')}$ and $a_{LIV}= \frac{\Delta E}{H_0
E_{QG}}$. The original paper \cite{Ellis2} contained an error in the formula defining $K(z)$ function. This was later corrected in \cite{Piran Errata} (correct expression was also given in \cite{Biesiada Piorkowska}), but apart from fixing an obvious mistake the correction only changed numerical values of the
results and was not able to erase the effect.

Such parametrization allows to formulate the problem in terms of linear regression:
\begin{equation} \label{regression}
\frac{\Delta t_{obs}}{1+z} = a_{LIV} K(z) + b
\end{equation}
where the intercept informs about intrinsic time lags, and slope carries information about LIV effects.

\section{Cosmological models tested}

One of the most important issues in modern cosmology is the problem of ``dark energy'', which appeared after the discovery of accelerated expansion of the Universe as inferred from the SNIa Hubble diagram \cite{Perlmutter}. Since then a lot of specific scenarios have been put forward as an explanation of this puzzling phenomenon. They fall into two broad categories: searching an explanation among hypothetical candidates for dark energy (cosmological constant $\Lambda$ \cite{Perlmutter}, quintessence - evolving scalar fields \cite{Ratra}, Chaplygin gas \cite{Kam}) or modification of gravity theory (supergravity
\cite{Brax}, brane world scenarios \cite{DGP}).

We will restrict our attention to flat models $k=0$ because the flat FRW geometry is strongly supported by cosmic microwave background radiation (CMBR) data \cite{Spergel,Boomerang}. Friedman - Robertson - Walker model with non-vanishing cosmological constant and pressure-less matter including the dark part of it responsible for flat rotation curves of galaxies (the co called $\Lambda$CDM model) is a standard reference point in modern cosmology. Sometimes it is referred to as a concordance model since it fits rather well to independent data (such like CMBR data, LSS considerations, supernovae data). The cosmological constant suffers from the fine tuning problem (being constant, why does it start dominating at the present epoch?) and from the enormous discrepancy between facts and expectations (assuming that $\Lambda$ represents quantum-mechanical energy of the vacuum it should be 55 orders of magnitude larger than observed
\cite{Weinberg}).

Hence another popular explanation of the accelerating Universe is to assume the existence of a negative pressure component called dark energy. One can heuristically assume that this component is
described by hydrodynamical energy-momentum tensor with (effective) cosmic equation of state: $p = w \rho$ where $-1 < w < -1/3$ \cite{Chiba98}. In such case this component is called "quintessence". Confrontation with supernovae and CMBR data \cite{BeanMelchiorri} led to the constraint $w \leq -0.8$. This was further improved by combined analysis of SNIa and large scale structure considerations (see e.g. \cite{Melchiorri}) and from WMAP data on CMBR \cite{Spergel}. The ESSENCE supernova survey team \cite{Wood-Vasey} pinned down the equation of state parameter to the range $w = -1.07 \pm 0.09 (stat) \pm 0.12 (systematics)$.
The most recent estimate of $w = -0.969^{+0.059}_{-0.063} (stat) ^{+0.063}_{-0.066} (systematics)$ comes from the Union08 compilation \cite{Kowalski}. For the illustrative purposes we chose $w = -0.87$ as representing a quintessence model which is different enough from cosmological constant and still admissible
by the data.

If we think that the quintessence has its origins in the evolving scalar field, it would be natural to expect that $w$ coefficient should vary in time, i.e. $w = w(z).$ An arbitrary function $w(z)$
can be Taylor expanded. Then, bearing in mind that both SNIa surveys or strong gravitational lensing systems are able to probe the range of small and moderate redshifts it is sufficient to
explore first the linear order of this expansion. Such possibility, i.e. $w(z) = w_0 + w_1 z$ has been considered in the literature (e.g. \cite{Weller}). Fits to supernovae data performed in the literature suggest $w_0=-1.5$ and $w_1=2.1$ \cite{JainAlcanizDev} (which is consistent with fits given in \cite{BiesiadaJCAP}). Therefore we adapted these values as representative for this parametrization of the equation of state.

In the class of generalized Chaplygin gas models matter content of the Universe consists of pressure-less gas with energy density $\rho_m$ representing baryonic plus cold dark matter (CDM) and of the generalized Chaplygin gas with the equation  of state $p_{Ch} =- \frac{A}{{\rho_{Ch}}^{\alpha}}$ with $0\le \alpha \le
1$, representing dark energy responsible for acceleration of the Universe. Using the angular size statistics for extragalactic sources combined with SNIa data it was found in \cite{AlcanizLima}
that in the the $\Omega_m =0.3$ and $\Omega_{Ch}=0.7$ scenario best fitted values of model parameters are $A_0=0.83$ and $\alpha=1.$ respectively. Generalized Chaplygin gas models have been intensively studied in the literature \cite{Makler} and in particular they have been tested against supernovae data (e.g. \cite{BiesiadaGodlowski2005} and references therein). Conclusions from these fits are in agreement with the above mentioned values of parameters so we used them as representative of Chaplygin Gas models.

Brane-world scenarios assume that our four-dimensional spacetime is embedded into 5-dimensional space and gravity in 5-dimensions is governed by the usual 5-dimensional Einstein-Hilbert action.
The bulk metric induces a 4-dimensional metric on the brane. The brane induced gravity models \cite{DGP} have a 4-dimensional Einstein-Hilbert action on the brane calculated with induced metric. According to this picture, our 4-dimensional Universe is a surface (a brane) embedded into a higher dimensional bulk
space-time in which gravity propagates. As a consequence there exists a certain cross-over scale $r_c$ above which an observer will detect higher dimensional effects. Cosmological models in brane-world scenarios have been widely discussed in the literature \cite{Jain}. It has been shown in \cite{Jain} that flat brane-world Universe with $\Omega_m=0.3$ and $r_c = 1.4 \;H_0^{-1}$ is consistent with current SNIa and CMBR data. Note that in flat (i.e. $k=0.$) brane-world Universe the following relation is valid: $\Omega_{r_c} = \frac{1}{4}(1-\Omega_m)^2$. Futher research performed in \cite{Malcolm2005} based on SNLS
combined with SDSS disfavored flat brane-world models. More recent analysis by the same authors \cite{Malcolm2007} using also ESSENCE supernovae sample and CMB acoustic peaks lead to the conclusion
that flat brane-world scenario is only slightly disfavored, although inclusion of baryon acoustic oscillation peak would ruled it out. Despite this interesting debate we use flat brane-world scenario with $\Omega_m=0.3$ for illustration.

Because in what follows we will use model selection approach, it would be tempting --- indeed even more appropriate from the methodological point of view --- to take model parameters as free and estimate their values from the best fits. However, with a small sample and considerable parameter space, it is almost sure
that the values (e.g. $\Omega_m$, $w$, $w_0$, $w_1$, etc.) inferred would be inconsistent with general knowledge already acquired with alternative techniques on much better data and much larger samples (like SNe Ia, CMBR, LSS). The major strength of modern cosmology is in consistency, therefore we take fixed values (best fits to SN Ia data) of cosmological parameters specifying the models considered.

\section{Searching for LIV effects in different cosmological models}

The sample we used consists of 35 GRBs with known redshifts for which time lags between different energy channels have been assessed from the light curves by Ellis et al. \cite{Ellis2}. They are summarized in Table 1 of \cite{Ellis2}. The data are based on the results of BATSE (9), HETE (15) and Swift (11) experiments. The numbers in brackets indicate the number of bursts detected in a specific experiment.
Technical details can be found in \cite{Ellis2}. We took these data for the sake of comparability.

The dependence on cosmological model resides in the $K(z)$ term which was calculated for all models considered. Their expansion rates expressed through cosmological model parameters are shown in
Table I. The parameter values we used were described in previous section. The results of linear regression are summarized in Table II and displayed in Figure 1. One can see that the LIV effect (i.e.
non-zero slope) is visible in all models. For most of them it is significant at $2\sigma$ level and for the quintessence model with varying equation of state at $3\sigma$ level. One should note, however, that our regression is a straightforward one not supported with sophisticated tricks original authors made after
such straightforward fit in order to get their final estimate. Therefore, we do not attempt to convert slope factors into LIV energy scale estimate.

One can see that in all classes of alternative cosmological models the effect is similar, hence there is no indication that Ellis et al. result might be an artifact of assuming $\Lambda$CDM. Now, one can ask which cosmological model ``is the best'' in the sense of revealing a sought-for LIV signature i.e. the linear relation (\ref{regression}).

In order to compare different models -- the problem we encounter here -- one can use Akaike information-theoretical model selection criterion \cite{Akaike73}. In particular this criterion has become a standard diagnostic tool of regression models \cite{Burnham}. In cosmology it has first been applied by Liddle \cite{Liddle} and then was used in e.g. \cite{BiesiadaJCAP,Szydlowski,Szydlo new}.

Akaike criterion is based on Kullback-Leibler information $I(f,g)$ between two distributions $f(x)$ and $g(x)$. The intuitive meaning of $I(f,g)$ (also called K-L divergence) is the information lost when $g$ is used to approximate $f$. It is convenient to think that $f(x)$ denotes the true mechanism behind the data and $g(x|\theta)$ its approximating model (parametrized by $\theta$). Of course, K-L divergence cannot be assessed without prior knowledge of the true model $f(x)$ as well as parameters $\theta$ of the approximating model $g(x|\theta)$. However, given $f(x)$ and $g(x|\theta)$ there exists the ``best'' value of $\theta$ for which Kullback-Leibler divergence is minimized. The maximum likelihood estimator $\hat{\theta}$ of $\theta$ parameter is exactly this K-L ``best'' one.

The core result of Akaike was in showing that an approximately unbiased estimator of K-L divergence is $ln({\cal
L}(\hat{\theta}|data))- K$ where $\cal L$ is the likelihood function (more precisely its numerical maximum value - taken at $\hat{\theta}$) and $K$ is the number of estimable parameters ($\theta$) in approximating model $g(x|\theta)$. For historical reasons Akaike formulated this result in the following form:
\begin{equation} \label{AIC}
AIC = - 2 ln({\cal L}(\hat{\theta}|data))+ 2K
\end{equation}
which became known as Akaike information criterion. Heuristically one may think of it as of an estimator of K-L divergence between the model at hand $g(x|\theta)$ and an unknown true model $f(x)$ which generated the data. The first term measures goodness of model fit (or more precisely the lack thereof) and the second one
(competing with the first) measures model complexity (number of free parameters).

The AIC value for a single model is meaningless (simply because the true model $f(x)$ is unknown). What is useful, instead are the differences $\Delta_i := AIC_i - AIC_{min}$ calculated over the whole set of alternative candidate models $i=1,...,N$ where by $AIC_{min}$ we denoted $min\{AIC_i ; i=1,...,N\}$. Comparing several models, the one which minimizes AIC could be considered the best. Then the relative strength of evidence for each model can be calculated as the likelihood of the model given the data ${\cal L}(g_i|data) \propto exp(-\frac{1}{2}\Delta_i) $. Relative likelihoods of the models ${\cal L}(g_i|data$) normalized to unity are called Akaike weights $w_i$. In Bayesian language Akaike weight corresponds to the posterior probability of a model (under assumption of equal prior probabilities). The (relative) evidence
for the models can also be judged by the evidence ratios of model pairs $\frac{w_i}{w_j} =\frac{{\cal L}(g_i|data)}{{\cal L}(g_j|data)}$.

A very similar criterion was derived by Schwarz \cite{Schwarz} in a Bayesian context. It is known as the so called Bayesian information criterion (BIC) (\cite{Schwarz}):
\begin{equation} \label{BIC}
BIC = - 2 ln({\cal L}(\hat{\theta}|data))+ K ln(n)
\end{equation}
where $n$ is sample size and as previously $K$ denotes number of parameters. BIC is not an estimator of K-L divergence -- its derivation stems from estimating the marginal likelihood of the data (marginalized over parameters).
BIC does not take the full advantage offered by Bayesian techniques such as described e.g in Trotta and Kunz \cite{Trotta Kunz} (see also references therein).

At last, it should be noted that AIC may perform poorly if the sample size is not large enough ($n<40$ as a rule of thumb), which is the case of our study. In order to cope with such situations Sugiura \cite{Sugiura 1978} derived a variant of Akaike criterion called c-AIC, which is related with AIC in the following manner: $AIC_c = AIC + \frac{2K(K-1)}{n-K-1}$.

In our case i.e. simple linear (univariate) regression the distinction between AIC, c-AIC and BIC is purely formal. Technically the corresponding numerical values are shifted by a constant, which does not matter since there are the differences that count. If one used the freedom of cosmological model parameters entering $K(z)$ while performing the fit, clearly the c-AIC should be used with the sample size of $n=35$. In our specific case of LIV testing this is not an option. Hence the reported AIC analysis summarized in Table III is representative for all above mentioned model selection criteria (and due to simplicity also to the standard goodness of fit techniques). The likelihoods needed to calculate AIC were taken here as exponents of minus chi-square for the best fitted model. One can see that the quintessence model with varying equation of state gives the best fit in time delay vs. $K(z)$ regression. However the odds
against the other models are not big enough to call it preferred.

\section{Discussion and conclusions}

In this paper we performed fits of time delays (between different energy channels in a sample of GRBs) versus the $K(z)$ function (dependent on the cosmic expansion rate) in five cosmological models representative to different dark energy scenarios. The technique we used was taken over an original one of Ellis et al. \cite{Ellis2} and so was also the sample adopted. This was motivated by suggestion formulated in \cite{Ellis2} that the LIV effect shows up in the studied sample of gamma ray bursts. These authors have used the ``concordance'' model (i.e. flat $\Lambda$CDM with $\Omega_{\Lambda} = 0.3$) as representing the
cosmological background.

Despite the fact that such $\Lambda$CDM model is commonly used as a reference point in cosmological studies, there is not so obvious that this is the true solution of accelerating Universe enigma. See e.g. the most recent discussion of this issue by Linder \cite{Linder} (and references therein). There exist several classes of alternative dark energy scenarios, from which we have taken four most commonly considered.

Initially we suspected that the effect noticed in \cite{Ellis2} could be an artifact of incorrectly assuming $\Lambda$CDM in their fits whereas another model better represents the accelerating
Universe. Contrary to our expectations the effect does not get smaller in alternative models. In fact it is the highest in the model providing the best fit of time delay vs. $K(z)$ regression.

Curiously the quintessence model with varying equation of state is the one which gives the best fit. It is most probably a mere coincidence but it is this cosmological model which is by many
considered to correctly reflect the phenomenology of the dark energy. Dark energy is the solution of presently accelerating expansion of the Universe, and, as brilliantly noticed in \cite{Linder}, the only accelerating expansion phase we know (i.e. the inflation) has ended hence clearly had time variation and
dynamics.

Our approach was a very simple one, aimed mainly to raise the issue. However the results suggest that it would be interesting to go beyond taking compilation of \cite{Ellis2} and re-analyze for the LIV effects, assuming alternative cosmological models of an accelerating Universe, on the raw data (GRBs light curves in
different energy channels) and with all the sophistication used in the original study.

\section*{Acknowledgements}
This work was partly supported by the Polish Ministry of Science Grant no N N203 390034

\begin{table}
\caption{Expansion rates $H(z)$ in the models tested. The
quantities $\Omega_i$ represent fractions of critical density
currently contained in energy densities of respective components
(like clumped pressure-less matter, $\Lambda$, quintessence,
Chaplygin gas or brane effects).}

\begin{tabular}{|c|c|} \hline
Model&Cosmological expansion rate $H(z)$ (the Hubble function)\\
\hline
$\Lambda$CDM&$H^2(z) = H^2_0 \left[ \Omega_m \; (1+z)^3 + \Omega_{\Lambda} \right]$\\
Quintessence&$H^2(z) = H^2_0 \left[ \Omega_m \; (1+z)^3 +
\Omega_Q \; (1+z)^{3(1+w)} \right]
$\\
Var Quintessence&$H^2(z) = H^2_0 \left[ \Omega_m \; (1+z)^3 +
\Omega_Q \;
(1+z)^{3(1+w_0-w_1)}\;\exp(3 w_1 z) \right]$\\
Chaplygin Gas&$H(z)^2 = H_0^2 \left[ \Omega_{m} (1+z)^3 +
\Omega_{Ch} \left(A_0 + (1 - A_0)(1+z)^{3(1+ \alpha)}
\right)^{{1\over 1+\alpha}} \right]
$\\
Braneworld&$H(z)^2 = H_0^2 \left[ (\sqrt{ \Omega_{m} (1+z)^3 +
\Omega_{r_c} }
+ \sqrt{\Omega_{r_c}} )^2 \right]$\\
\hline
\end{tabular}
\end{table}

\begin{table}
\caption{Regression coefficients (with $1 \sigma$ ranges) for the
time delay vs. $K(z)$ technique in the cosmological models
tested.}

\begin{tabular}{|c|c|c|} \hline
Cosmological model&Regression coefficient $a_{LIV}$&Intercept $b$\\
\hline
$\Lambda$CDM&$a_{LIV}= -0.0794 \pm 0.0447$&$b = 0.0494 \pm 0.0288$\\
Quintessence&$a_{LIV}= -0.0806 \pm 0.0460$&$b = 0.0489 \pm 0.0288$\\
Var Quintessence&$a_{LIV}= -0.1510 \pm 0.0683$&$b = 0.0735 \pm 0.0340$\\
Chaplygin Gas&$a_{LIV}= -0.1201 \pm 0.0618$&$b = 0.0627 \pm 0.0330$\\
Braneworld&$a_{LIV}= -0.0866 \pm 0.0493$&$b = 0.0501 \pm 0.0294$\\
\hline
\end{tabular}
\end{table}

\begin{figure}[htp]
\centering
\caption{Results of the linear regression for time delay vs K(z) technique in the cosmological models tested (the case of Quintessence was too similar to the $\Lambda$CDM to deserve displaying).}
\centering
\includegraphics{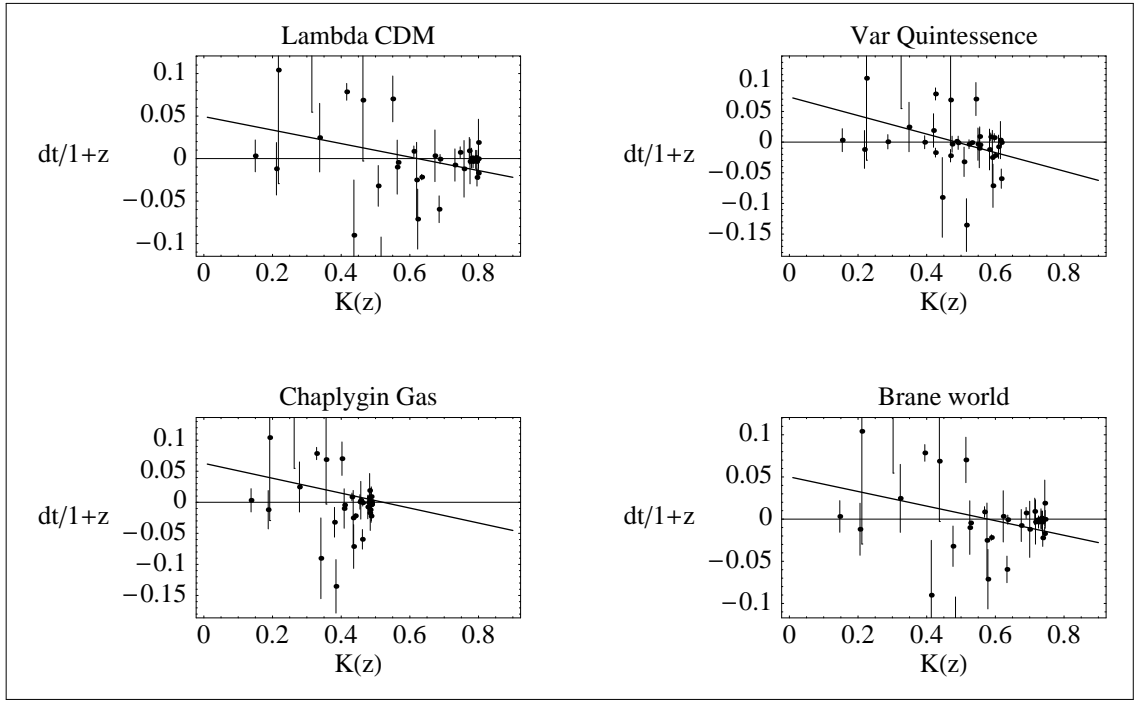}
\end{figure}

\begin{table}
\caption{Values of AIC, Akaike differences, Akaike weights $w_i$
(in Bayesian language equivalent to posterior model probabilities)
and odds against the model (with respect to the best fitted one).}

\begin{tabular}{|c|c|c|c|} \hline
Model&$\Delta_i$& $w_i$&Odds against\\
\hline
$\Lambda$CDM&1.645& 0.152& 2.276\\
Quintessence&1.712&0.147&2.354\\
Var Quintessence&0.&0.347&1.\\
Chaplygin Gas&1.042&0.206&1.684\\
Braneworld&1.704&0.148&2.344\\
\hline
\end{tabular}
\end{table}

\end{document}